\documentclass[twocolumn]{aastex631}

\usepackage{color,soul}
\usepackage{graphicx,float}
\usepackage{comment}

\begin{document}
\title{Temperature Diagnostics of Chromospheric Fibrils using DKIST/ViSP Observations: K-Means Clustering Approach}

\author[0000-0002-5504-6773]{Sanjay Gosain}
\affiliation{National Solar Observatory \\
3665 Discovery Dr., \\
Boulder, CO 80303, USA}

\begin{abstract}
The chromosphere is a critical layer of the solar atmosphere situated between the photosphere and the corona. Studying its temperature structure is important to understand the complex dynamics and energy-transfer processes between these layers. We investigate the thermodynamic properties of chromospheric fibrils adjacent to a plage region using high-resolution DKIST/ViSP observations of the Ca II 854.2 nm spectral line. We analyze the spectral profiles with the non-LTE inversion code NICOLE combined with K-means clustering. The high spectral and spatial resolution of the DKIST observations allows us to trace thermodynamic properties—temperature, density, line-of-sight velocity, and microturbulent velocity—along and across the fibrils.  We note that while the thermodynamic parameters are retrieved under the assumption of hydrostatic equilibrium, the resulting density and temperature values should be interpreted with the caveat that dynamic and magnetic terms are neglected. The temperature along the fibril length decreases smoothly by about 1000 K from the hotter footpoints toward the mid-axis. The temperature variation across the lateral boundary of fibrils is more abrupt and can vary by several hundreds of degree Kelvin across a megameter. Denser fibrils tend to be associated with cooler, downflowing plasma, while less-dense fibrils do not show this trend. Furthermore, the hotter parts of the fibrils tend to exhibit higher microturbulent velocities than the cooler parts.

\end{abstract}
\keywords{Sun--Chromosphere; Sun--Magnetic Fields; Sun--Polarimetry}

\section{Introduction} \label{sec:intro} 
Chromospheric fibrils are thin, elongated and dark features observed in the monochromatic images taken in the chromospheric spectral lines such as H$\alpha$, Ca II 854.2 nm, and Ca II H\&K. They appear prominently on the periphery of magnetic flux concentrations such as plages, enhanced magnetic network and active regions. As the photospheric magnetic flux expands into the chromosphere due to the decreasing gas pressure, the magnetic flux bundles arch over and become nearly horizontal forming a magnetic canopy that we see as fibrils and can extend several thousand kilometres, i.e., supergranular scales \citep{leenaarts2012}.

Polarimetric observations of the Ca~II~854.2~nm and He~I~1083~nm lines reveal that the fibrils are not perfectly aligned with magnetic field and misalignments of 5--20$^{\circ}$ are common \citep{delacruz2011,schad2013,aasensio2017}. \cite{Sykora2016} argued from numerical simulations that such misalignment can arise due to ambipolar diffusion in the partially ionized chromospheric plasma. State-of-the-art numerical simulations of the solar chromosphere \citep{gudiksen2011,rempel2017} have been used to produce synthetic observables reproducing such structures \citep{leenaarts2012,Bjorgen2019}. These simulations show that fibril-like structures naturally arise as ridges of enhanced mass density and cooling in the dynamic chromosphere. 

Another class of short and dynamic fibrils (DF), observed everywhere on the disk, not just strong flux concentrations, are the spicules. These are more vertical and dynamic and have shorter lifetimes as compare to horizontal fibrils. These dynamic fibrils appear in two flavors, so-called type-I and type-II spicules, and are believed to be the conduits of mass and energy transfer between the dense and dynamic photosphere and the tenuous hot corona, mostly via jets and magnetohydrodynamic (MHD) waves \citep{Zaqarashvili2009}.   

In this work, however, we are mainly concerned with the long horizontal fibrils present in the canopy adjacent to an enhanced network. Characterizing the thermodynamics of these features using high-resolution observations is important for understanding the interlaced structure of bright and dark fibrils and their thermal and density stratification.   

\cite{Kriginsky2023} studied fibrils using high-resolution observations from the Swedish Solar Telescope (SST) and found that the temperature of fibrils varies along their length, with the footpoints being about 300 K hotter than the midpoints on average. Microturbulent velocities showed a similar behavior to the temperature. They also reported that the temperatures of dark fibrils were lower than those derived from cloud-model calculations.

Here we analyze high-resolution spectroscopic observations of a plage region and adjacent fibrils obtained in the Ca II 854.2 nm spectral line with the Visible Spectro-Polarimeter (ViSP) instrument \citep{dewijn2022} at the Daniel K. Inouye Solar Telescope (DKIST) \citep{rimelle2020}. The Ca II 854.2 nm line forms in the lower–middle chromosphere and is an excellent diagnostic of this region of the solar atmosphere \citep{quintero2016} and whose interpretation requires non-local thermodynamic equilibrium (NLTE) radiative transfer treatment \citep{navarro2015,delacruz2017}.

NLTE inversion methods are computationally expensive and often require massively parallel computation to analyze observations in a reasonable timescale \citep{navarro2015,delacruz2019}. These computations are expensive because they require iteratively solving for the radiation field and atomic level populations in a self-consistent manner, accounting for multi-directional, multi-frequency coupling throughout the atmosphere \citep{deltoro2016}.

An important criterion when applying inversion methods is to start with a good initial guess model atmosphere, which typically helps to converge to a best-fit solution faster. Different approaches exist to find a good initial model for initialization such as use of neural networks, principal component analysis (PCA) and k-means clustering of spectra \citep{aasensio2017,dalda2019,milic2020,Gafeira2021,Navarro2021}. 

Here we follow the approach of \cite{dalda2019}, which uses k-means clustering of the observed spectra. Such clustering reduces the dimensionality of the data, allowing the observed region to be represented by a limited number of spectral profiles. The model atmosphere derived from the inversion of each cluster profile is used to initialize the pixel-by-pixel inversion of the full observed region. This improves convergence, smoothness and overall speed of the inversions, especially for large datasets.

In the following sections we first describe the observations and the method of data reduction in Section~2. The k-means clustering and subsequent NLTE analysis of the spectra are described in Section~3. The results pertaining to the retrieved model atmospheres, i.e., the thermodynamic characteristics of the fibrils and the neighboring plage region, are described in Section~4. Finally, we discuss the results and present our conclusions.

\begin{figure}[h]
\centering
\includegraphics[width=0.5\textwidth]{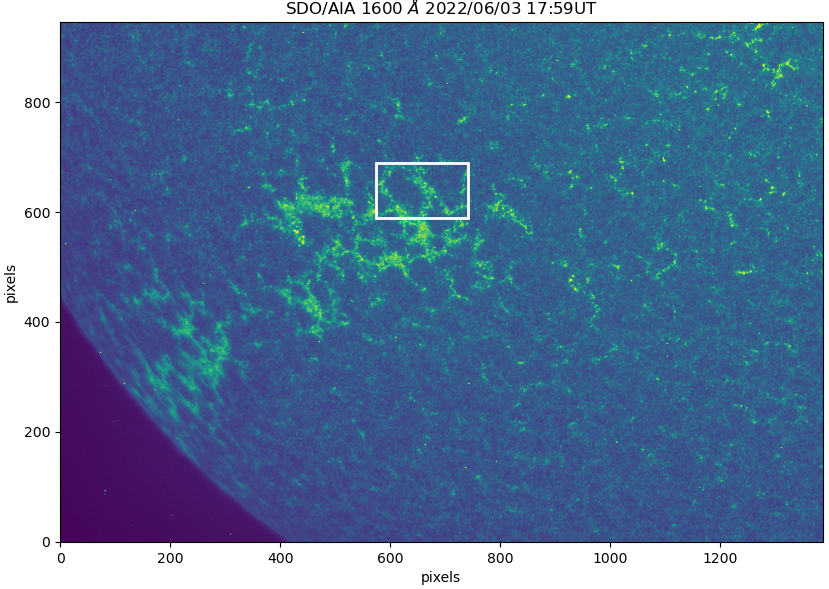}
\includegraphics[width=0.5\textwidth]{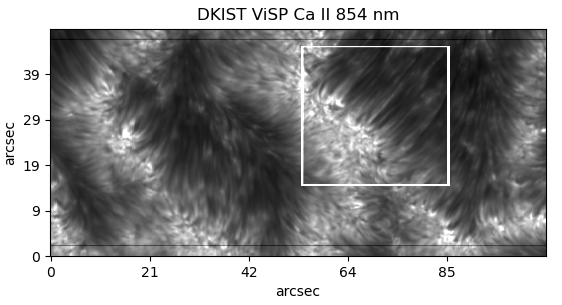}
\caption{The top panel marks the region scanned by the ViSP instrument over a context image of the Sun by SDO AIA in 1600\AA. The bottom panel shows the map of Ca II 854 nm line core intensity and a sub-region marked by a white rectangle, which is chosen for further analysis.  }
\label{fig:fig_roi}
\end{figure}

\begin{figure*}[t]
\centering
\includegraphics[width=0.44\textwidth]{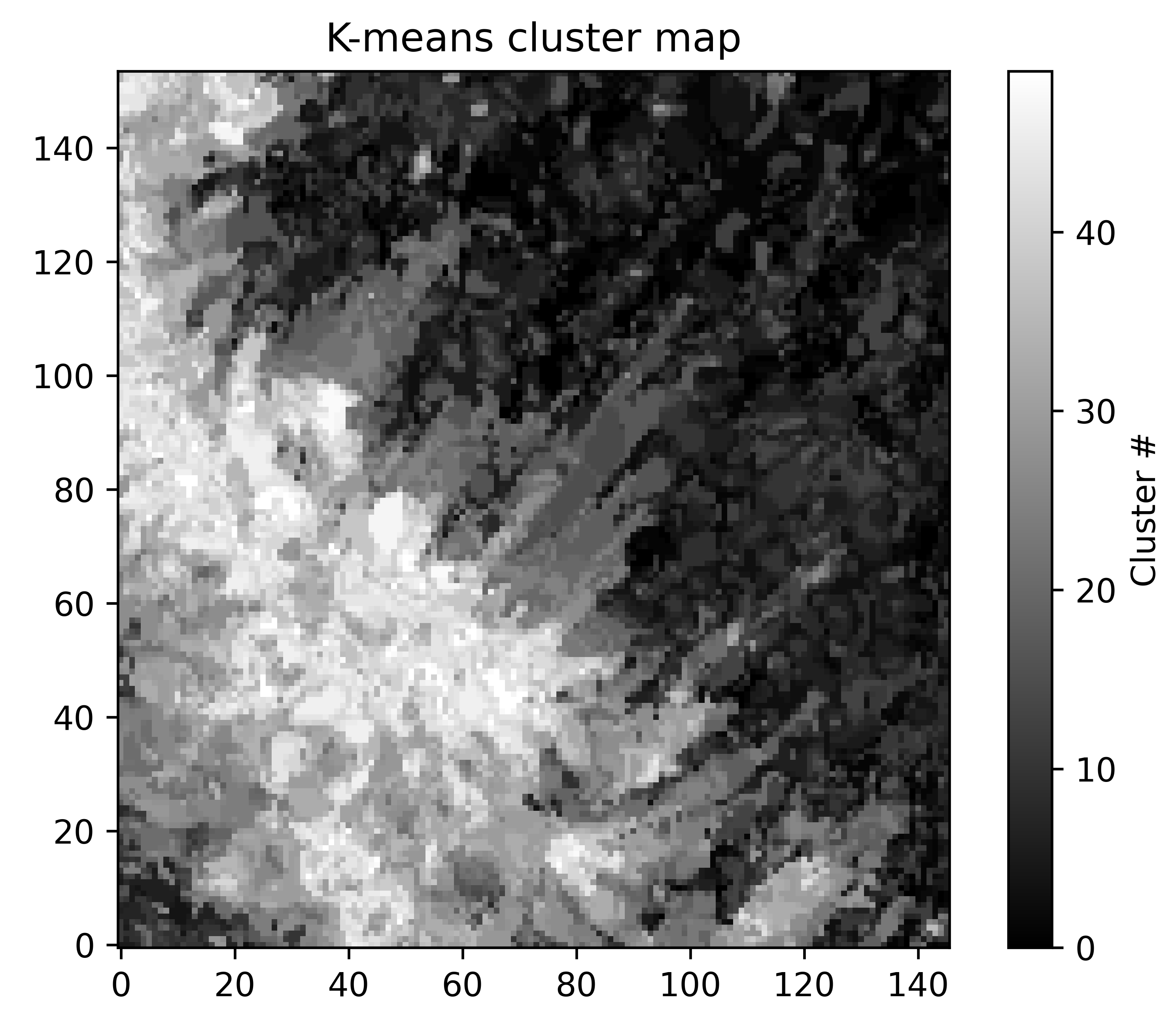}
\includegraphics[width=0.53\textwidth]{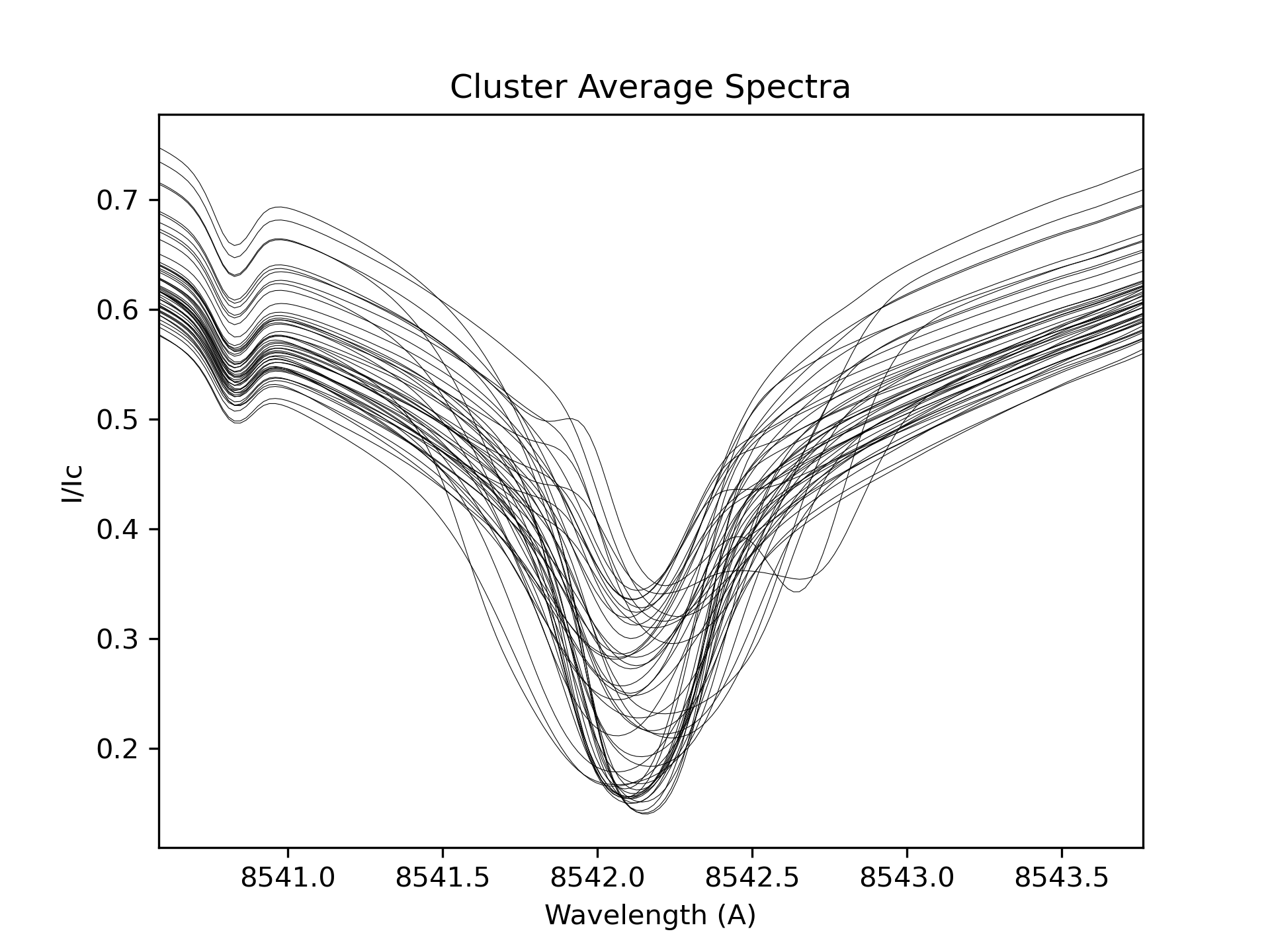}
\caption{ The left panel shows a map of cluster labels, color-coded in grayscale (a total of 50 clusters), corresponding to the observed region shown by white rectangle in the lower panel of Figure~\ref{fig:fig_roi}. The average profiles for each of the cluster is shown in the right panel.}
\label{fig:kmeans_clustermap}
\end{figure*}

\subsection{Observations and Data Reduction}
The observations analyzed here were obtained with the ViSP instrument \citep{dewijn2022} at the DKIST facility \citep{rimelle2020}. These observations are part of the first level-1 DKIST dataset released to the public (program ID 1.118). The data were obtained on 2022 June 3, 17:38--18:05 UT at heliocentric coordinates (-360", -410"). The spatial sampling along the slit is estimated to be 0.0194" per pixel; the slit scan step is 0.219" and the slit width is 0.214" \citep{santos2023}.

The nominal dispersion is 1.882 pm/pixel at 854 nm, with a ~1\% change over the observed bandwidth. The spectral calibration provided in the level-1 data is achieved by fitting the positions of the strongest telluric lines in the spectrum using the NSO/FTS Telluric Atlas. A quadratic polynomial is then fit to these values. The spectral sampling is found to be 1.884 pm/pixel, in good agreement with the nominal value. The spectral resolution of ViSP at 854 nm is estimated to be about 120,000 \citep{santos2023}. The observed spectral range spans from 853.7 nm to 854.6 nm, covering the full Ca II 854.2 nm line profile. The exposure time per slit position is 100 ms, and the scan consists of 200 steps, resulting in a total scan time of approximately 27 minutes. The data have been dark-subtracted, flat-fielded, and corrected for instrumental polarization using the DKIST Data Handling System (DHS) pipeline.

The second step in spectral calibration is continuum normalization, i.e., normalizing the observed spectrum to the quiet-Sun disk-center continuum. The solar region analyzed here corresponds to a heliocentric angle of $\mu\sim0.85$. Due to the limited FoV of ViSP, simultaneous disk-center observations are not possible. Hence one typically uses the NSO/FTS atlas spectrum \citep{Kurucz1984} together with the relevant limb-darkening function \citep{Neckel2005} for spectrum normalization. We follow a similar approach; however, we use full-disk Ca II 854 nm observations from the SOLIS/VSM telescope \citep{keller1998} to determine the CLV correction (see Appendix A).

In this work we use only the Stokes-I profiles of the Ca II 854.2 nm line, because we focus on the thermodynamic properties of the observed features. The top panel of Figure~\ref{fig:fig_roi} shows the location of the ViSP scan (white box) overlaid on an SDO/AIA 1600\AA\/ image. The bottom panel shows a map of the Ca II 854 nm line-core intensity; the region marked by the white box is analyzed in more detail. This region was selected because it samples a bright plage region in the lower left and dark fibrils in the upper right, allowing us to diagnose thermodynamic characteristics of diverse features. The fibrils in this portion of the field of view (FoV) are relatively uncomplicated (with less overlap) and appear well separated along their elongated axes.

\begin{figure*}[t]
\centering
\includegraphics[width=0.99\textwidth]{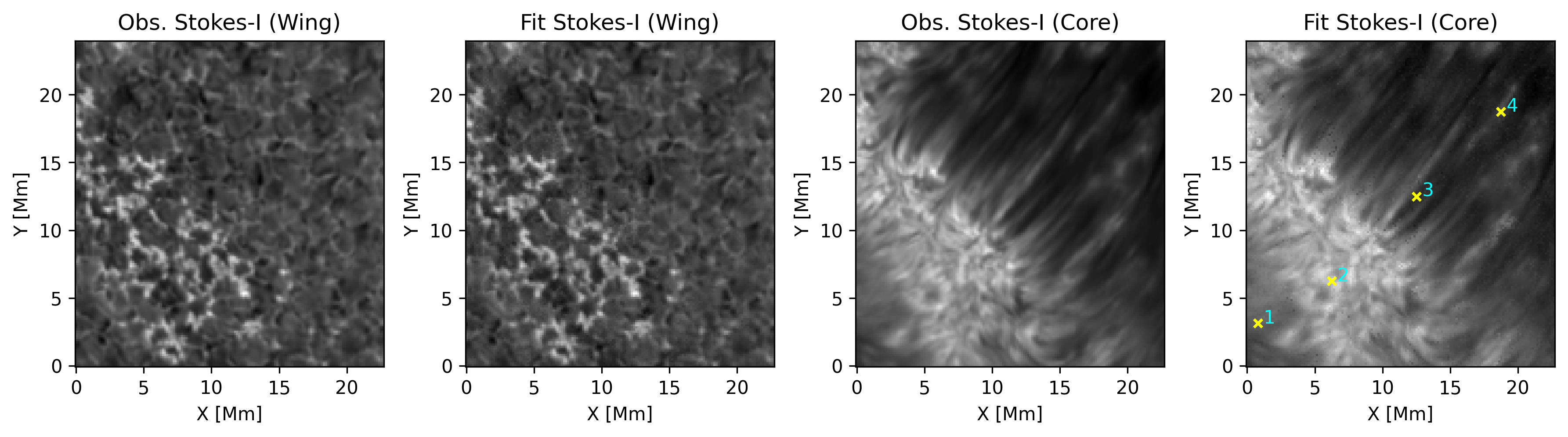}
\includegraphics[width=0.99\textwidth]{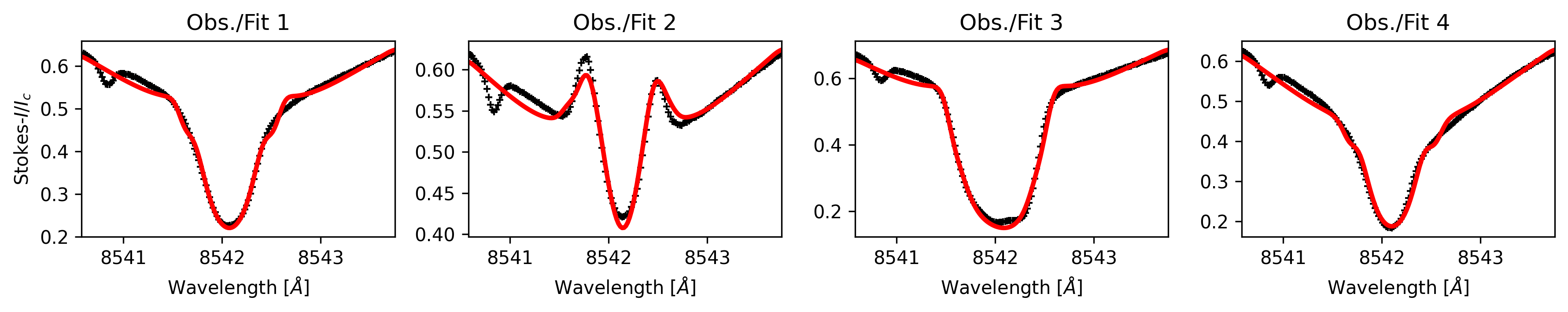}
\includegraphics[width=0.99\textwidth]{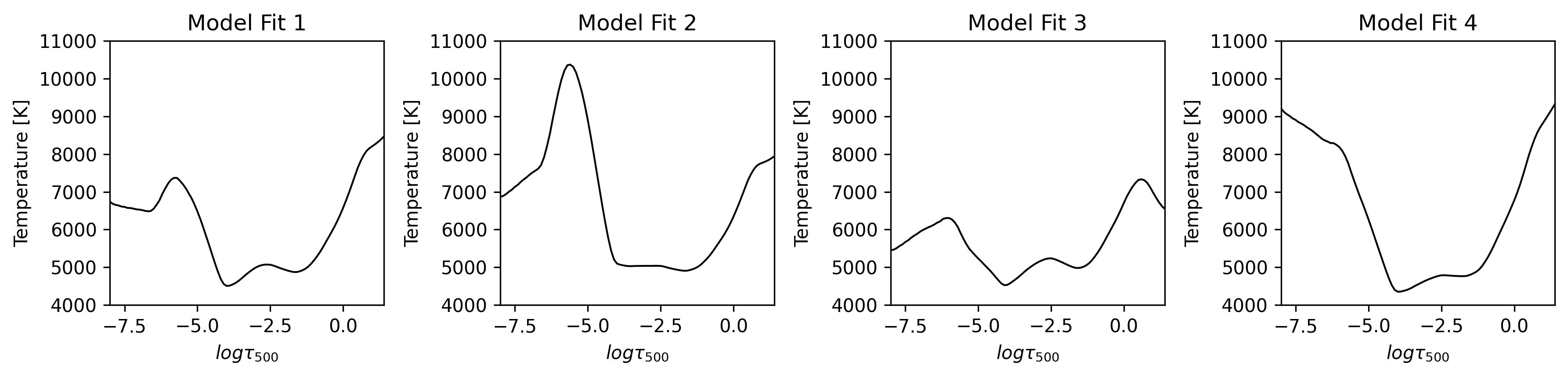}
\caption{The top row shows the maps of Ca II line core/wing intensity of the observed and the best-fit spectra, as indicated by the title on top of each panel, respectively. Four locations marked by 'x' symbols and labeled '1' through '4' on the top right panel, are chosen to show individual observed (black symbols)  and best-fit (red curve) spectral profiles. The bottom row shows the models of temperature  versus optical depth retrieved by fitting the observed spectral profiles, as shown in the middle row. }
\label{fig:fits_core_wing}
\end{figure*}

\subsection{Data Analysis}
The physical parameters in the line-forming region are inferred using the non-LTE inversion code NICOLE \citep{navarro2000}. The code seeks a model atmosphere that provides the best fit (in the least-squares sense) to the observed profiles. NICOLE assumes atomic populations in statistical equilibrium and complete angle and frequency redistribution, approximations that have been shown to be valid for the Ca II 854 nm line \citep{Uitenbroek1989,Wedemeyer2011}. The code combines SVD techniques with the Levenberg–Marquardt minimization method to solve the inverse problem \citep{Press1992}.

NLTE inversions are slow primarily due to the complex physics involved in line formation and the need for iterative solutions to statistical equilibrium and radiative transfer equations. Using a good guess for the model atmosphere therefore helps in the convergence of the iterative solutions. Here we use an approach similar to \cite{Pietarila2007,dalda2019} where they used K-means clustering of the spectral profiles observed over variety of solar features. The main idea behind this approach is that any observed region has many spectral profiles which are very similar in shape and originate from similar features on the Sun and therefore, can be initialized with the same guess model atmosphere for inversions.

\subsubsection{K-means Clustering}
K-means clustering is an unsupervised machine learning technique used for identifying hidden patterns or clusters within large datasets by grouping similar data points into clusters \citep{macqueen1967,lloyd1982}. This method is particularly effective in applications involving large volumes of data, such as the analysis of thousands of solar spectra across the observed field-of-view (FoV). When applied to our spectral data, this method groups spectra originating in similar features such as fibrils, networks, plage etc. or neighboring locations with similar spectral shapes into distinct clusters. While statistical diagnostics like the Elbow Method \citep{thorndike1953} suggested an optimal clustering at $k \approx 15$, we adopted a more granular value of $k=50$ to prioritize the detection of physically significant, low-population spectral features. Our analysis confirmed that at $k=50$, the algorithm successfully isolates rare but spatially coherent structures which exhibit distinct line-core intensities and thermal signatures that would otherwise be lost to averaging in a lower-order model. Furthermore, the Doppler shift distribution across these 50 clusters reveals a high degree of dynamical diversity, capturing localized high-velocity flows. By maintaining this higher cluster count, we ensure that the k-means labels act as a sensitive proxy for both the thermal and kinetic heterogeneity of the solar atmosphere, providing a more robust foundation for identifying small-scale solar phenomena within the DKIST/VISP dataset.

As shown in figure~\ref{fig:kmeans_clustermap}, the panel on the left shows the map of 50 clusters, coded in grayscale such that the cluster index number is sorted according to the line core intensity. It can be noticed that the neighboring regions in various structures are grouped as one cluster. We can notice that spectra from quiet Sun, fibrils, plage are grouped in separate clusters. The spectrum averaged over each cluster is shown in the right panel of figure~\ref{fig:kmeans_clustermap}.

\begin{figure*}[t]
\centering
\includegraphics[width=1.\textwidth]{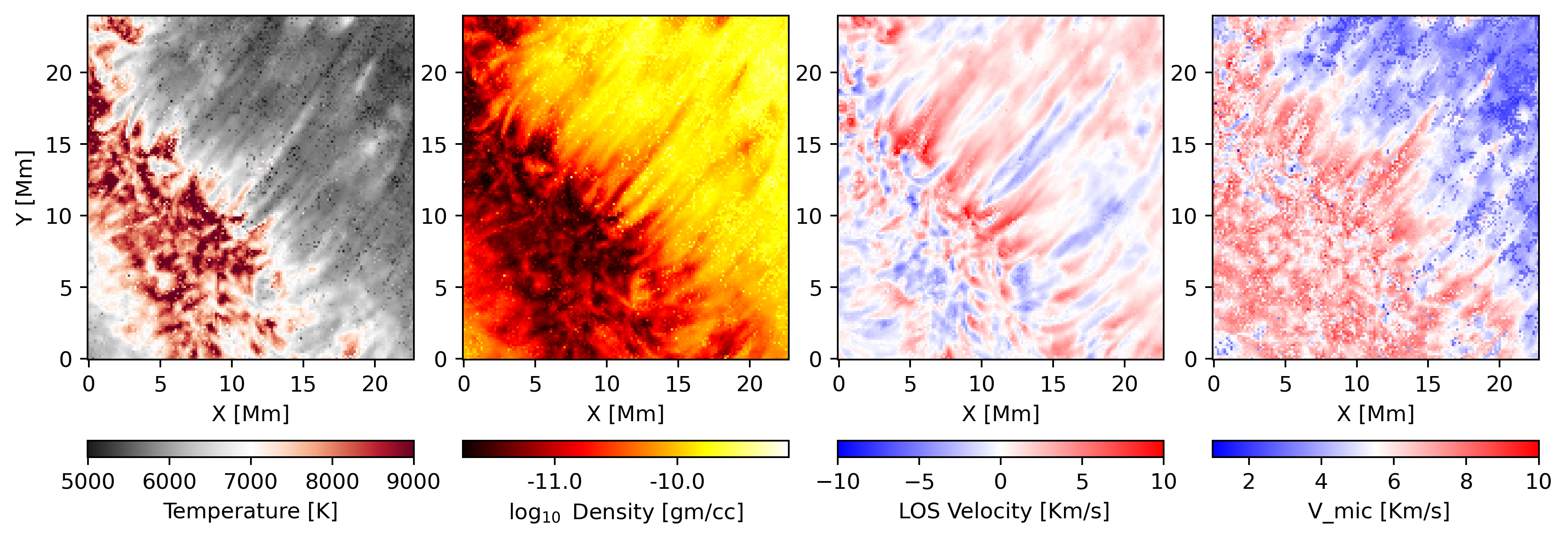}
\includegraphics[width=1.\textwidth]{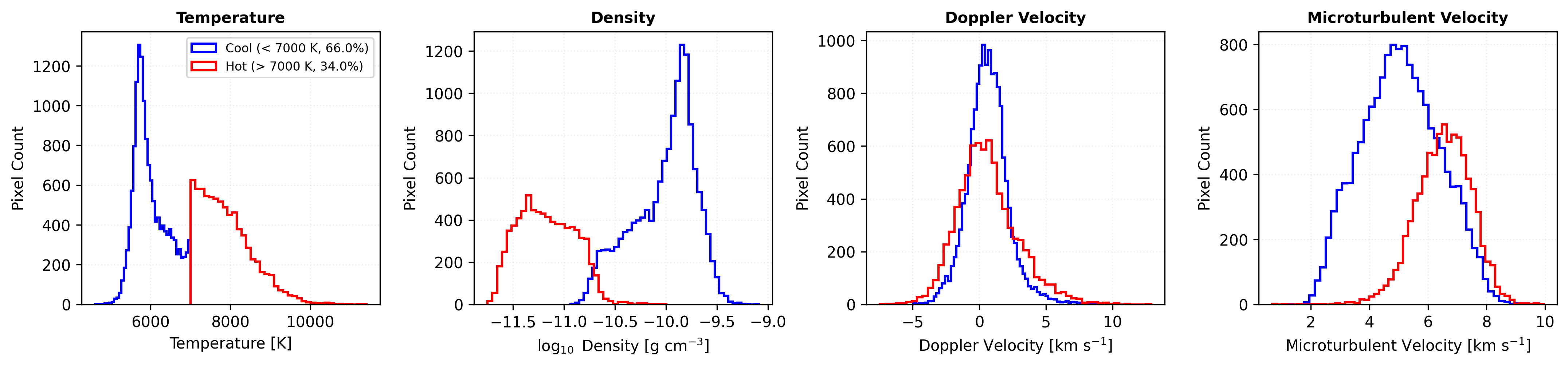}
\caption{The inverted maps of temperature, gas density, Doppler velocity and microturbulent velocity, averaged over an optical depth range of log$\tau$-4.5 to -5.5, are shown from left to right in the top row, respectively. Bottom row shows histograms of the corresponding maps.}
\label{fig:invmaps_temp_dens_dopp}
\end{figure*}

\begin{table}[t!]
\centering
\caption{Number of nodes used in the inversions.}
\label{tab:nodes}
\begin{tabular}{lccccc}
\hline \hline
Inversion & $T$ & $v_{\text{LOS}}$ & $v_{\text{turb}}$ & $N_{\text{inv}}$ & $N_{\text{cycle}}$ \\ \hline
Cluster-Averaged & 8 & 4 & 2 & 20 & 1   \\
Pixel-by-pixel   & 8 & 4 & 1 & 10  & 1  \\  \hline
\end{tabular}
\end{table}

\subsubsection{NLTE Inversions}
Our strategy for the full map NLTE inversion is two-step process, first invert the cluster-averaged spectra and get a best-fit model atmosphere for each cluster, then use these atmospheres to initialize the pixel-by-pixel inversion of the entire map. In both steps we use the NLTE inversion code NICOLE \citep{navarro2000}. 

In the first step, the cluster-averaged spectra, which we call representative profiles (RP), are inverted and the best-fit model atmospheres retrieved are called representative model atmosphere (RMA), respectively (this is the nomenclature used by \citep{dalda2019}). The guess model atmosphere used for these first step inversions was the Harvard Smithsonian Reference Atmosphere (HSRA; \cite{gingerich1971}) which ranges from optical depth log~$\tau$= 1.5 to -8.0 and contains no magnetic field.    

In the second step, we perform pixel-by-pixel inversion of the entire map where we initialize each cluster with its RMA  as a starting guess model. Such initialization with RMA is more robust since the spectral profiles in a cluster are spatially coherent and have similar shapes and thus are presumably well represented by a RMA \citep{dalda2019}. 

The number of nodes (free parameters) used in the inversion for temperature, line-of-sight velocity and microturbulent velocity are given in the Table~\ref{tab:nodes}. The second-last column specifies the number of times the inversion is repeated, each time with a small perturbation to the initial guess model, to reach a robust chi-square.  For cluster-averaged spectra we run 20 inversions to allow the code to find the best possible solution, given the initial guess model. On the other hand for the pixel-by-pixel inversion, only 10 inversions were sufficient as we start with a guess model atmosphere which is already close to the potential best-fit solution and also helps save the computation time. Further, as indicated in the last column, only one cycle of inversions was enough with no further change in number of nodes required.

The quality of fits is much better with our two-step inversion process as demonstrated by tests with simple one-step inversion process where HSRA atmosphere is used to initialize the full pixel-by-pixel inversions. The results of these tests are given in more detail in the Appendix-B.

\subsubsection{Best-fit spectral profiles versus the observations}
The quality of the spectral fits is shown in the top row of figure~\ref{fig:fits_core_wing}, where the panels from left to right show the maps of fitted and observed intensity in the wings and core of the spectral line, respectively. A good correspondence can be noted between the synthetic and observed maps in the wing and core of the line. Further, we show a few examples of the observed and fitted spectra at specific locations, marked by yellow asterisks on the top right panel, as shown in the bottom row panels. The black colored symbols are the observations and the red curve is the best-fit synthetic profile. These four locations are selected to represent diverse profiles representing quiet Sun, plage, fibril near plage region with blue-shift and a fibril far away from the plage region. The fitted profiles show good agreement with the observations. In summary, these best-fit spectra demonstrate the usefulness of RMA initialization approach as described the previous subsection.

\subsection{Results}
\subsubsection{Maps of Thermodynamic Parameters}
The  maps of temperature, gas density, line-of-sight (LOS) velocity, and microturbulent velocity derived from the best-fit model atmosphere are shown in the upper row of Figure~\ref{fig:invmaps_temp_dens_dopp}, while their distribution is shown in the corresponding histograms in the bottom row.  These parameters are averaged over the optical depths corresponding to $\log\tau = -4.5$ to $-5.5$, which is the height range where the core of the Ca II 8542~\AA line typically forms \citep{quintero2016}. These maps clearly show that the chromospheric fibrils are thermodynamically very distinct from their surroundings. For example, in the temperature map, fibrils appear as elongated cool structures in stark contrast to the adjacent hotter plage regions. 

\textit{Temperature Maps: }
Based on visual appearance of temperature map we separate the pixels into areas where the temperature is less than or greater than 7000 K. This temperature is represented by white color on the temperature colorbar and dilineates the colorscale into two halves. The histograms panels in the bottom row of the Figure~\ref{fig:invmaps_temp_dens_dopp} represents these areas by blue ($<$7000K) and red ($>$7000K) colored curves, respectively. By doing so we can study the distribution of other physical parameters, i.e., density, LOS velocity and microturbulent velocity in these distinct hotter and cooler regions. 

In the temperature histogram one can notice blue curve with a tall and narrow peak near 5800 K representing modest temperature of the cool fibrils, while their temperature spectrum spans 5000K to 7000K.  On the other hand, the hotter plage region shows broader range of temperatures ranging from 7000 to 10000 K, with a monotonous decrease in the area occupied by the hotter regions. The boundary between the plage and fibrils though arbitrarily defined around 7000 K (white colored region in the temperature map), the histogram value at this temperature shows that these regions are more plage-like rather than cool fibrils.

\textit{Density Maps: }
The map of density in the Figure~\ref{fig:invmaps_temp_dens_dopp} refers to the mean of the logarithm of the density, $\langle \log_{10} \rho \rangle$, over the optical depth range $\log\tau = -4.5$ to $-5.5$.
The density map shows enhanced density along many fibrils, indicating that these dark structures are generally mass-loaded relative to the plage chromosphere. The density in fibrils is about an order of magnitude larger than in the plage region. 

In the histogram panel for density, we notice that the cool and hot atmospheres have distinct density distributions, the peaks of which are separated by more than an order of magnitude. 
Similar to temperature the distribution of fibril density is much narrower than the density of hot plage regions, suggesting that fibrils have more homogeneous atmosphere than the plage, at least in the fibrils whose density exceeds $10^{-10}$ gm/cc. 

 It should, however, be noted that the gas density is derived under the assumption of hydrostatic equilibrium (HSE) which introduces specific caveats for the inferred semi-empirical models. While HSE primarily constraints the pressure and density scale, it also indirectly influences the retrieved temperature stratifications. In the non-LTE regime, the inversion code fits the observed line profiles by adjusting the source function, which remains coupled to both local temperature and density (see Section 4.2 of \cite{henriques2020}). Furthermore, the breakdown of HSE in the solar chromosphere is driven not only by the time-dependent dynamical effects but also by terms omitted from the simplified momentum equation. These include the velocity advective term and the Lorentz force, which may play a significant role in the force balance of magnetic fibrils.

Indeed, high-sensitivity spectropolarimetric analysis of this identical dataset by \cite{santos2023} revealed clear chromospheric Stokes $Q$, $U$, and $V$ signatures within the fibril canopy, corresponding to total magnetic field strengths of 200–300 G. At these field strengths the Lorentz force is expected to dominate the gas pressure gradients ($\beta \lesssim 1$), specially at the chromospheric heights \citep{gary2001}. Addressing this limitation in future studies will require moving beyond 1D HSE approximations (e.g., \cite{vincente2026}).

\textit{LOS Velocity Maps: }
The LOS velocity map shows a mixture of upflows and downflows, but redshifts dominate near the footpoints of fibrils rooted near the edge of the plage region. This increase in redshift is partially due to the projection effects, i.e., as the field lines that trace the horizontal fibrils bend towards the photosphere near the footpoints, the plasma downflow along the legs of the fibrils is better aligned to the line-of-sight, as illustrated by \cite{kuridze2024} (see, their cartoon in Figure 11). In addition to general downflow along fibrils we also notice few fibrils that exhibit blueshifts along their axis. 

The histogram of velocity shows that both hot and cool regions exhibit both red and blue shifts within the range of $\sim 5$--$7$~km\,s$^{-1}$ amplitude. The blue curve is however slightly red-shifted relative to the red curve suggesting a net downflow along the fibrils.   

\textit{Microturbulent Velocity Maps:}
The microturbulent velocity map shows enhanced non-thermal broadening in fibrils, with $v_{\rm mic}$ commonly reaching $\sim 5$--$7$~km\,s$^{-1}$. This is true of the portion of fibril lengths which are closer to the plage region , while further away the values drop to  $\sim 2.5$--$4.5$~km\,s$^{-1}$.  In contrast, the plage region shows enhanced microturbulent velocity $\sim$$7$~km\,s$^{-1}$ pretty much everywhere.

The histogram of microturbulent velocity clearly shows that the hotter regions (red curve) have significantly higher values, centered around $\sim$$7$~km\,s$^{-1}$, ranging from $\sim 4$--$9$~km\,s$^{-1}$. While cooler regions (blue curve) peaks at lower value $\sim$$5$~km\,s$^{-1}$, it spans a much larger range $\sim 2$--$9$~km\,s$^{-1}$ than the hotter regions alone. This is easily interpreted from the map of microturbulence where we can notice that the fibrils closer to the plage region exhibit much stronger microturbulent velocity which decreases rapidly along their axis away from the plage region. In summary, these maps indicate that the portion of the fibrils near the plages correspond to cool, dense and dynamically active plasma structures.

\begin{figure}[t]
\centering
\includegraphics[width=0.4\textwidth]{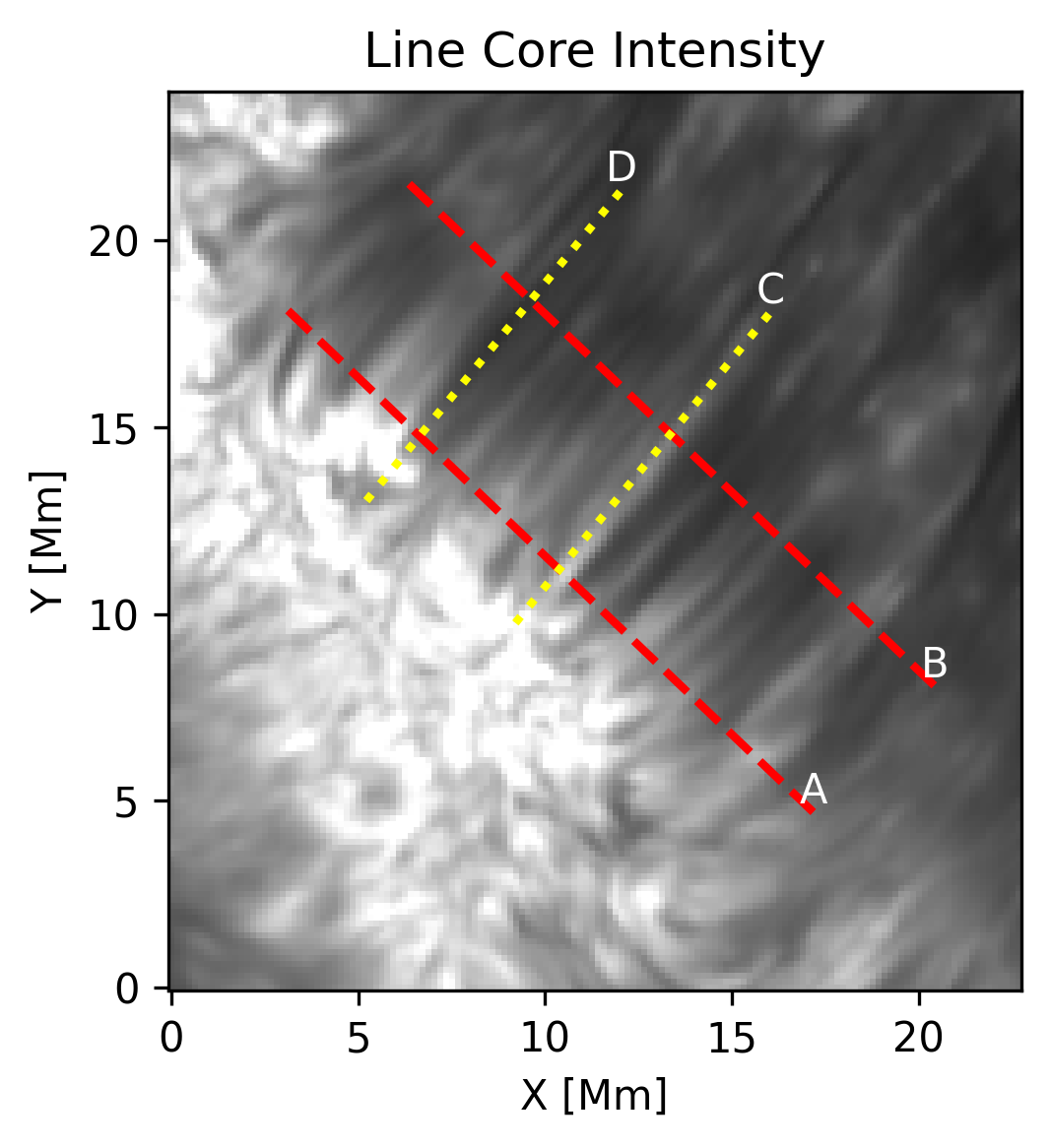}
\caption{Map of Ca II line core intensity is shown overlaid with four dashed/dotted lines which mark the location along which we sample the profile of physical parameters. These dashed (red)  and dotted (yellow)  lines are oriented such that they sample the fibrils across and along their long axis, respectively. These cuts are labeled alphabetically for reference.}
\label{fig:map_with_cuts}
\end{figure}

\begin{figure*}[t]
\centering
\includegraphics[width=0.465\textwidth]{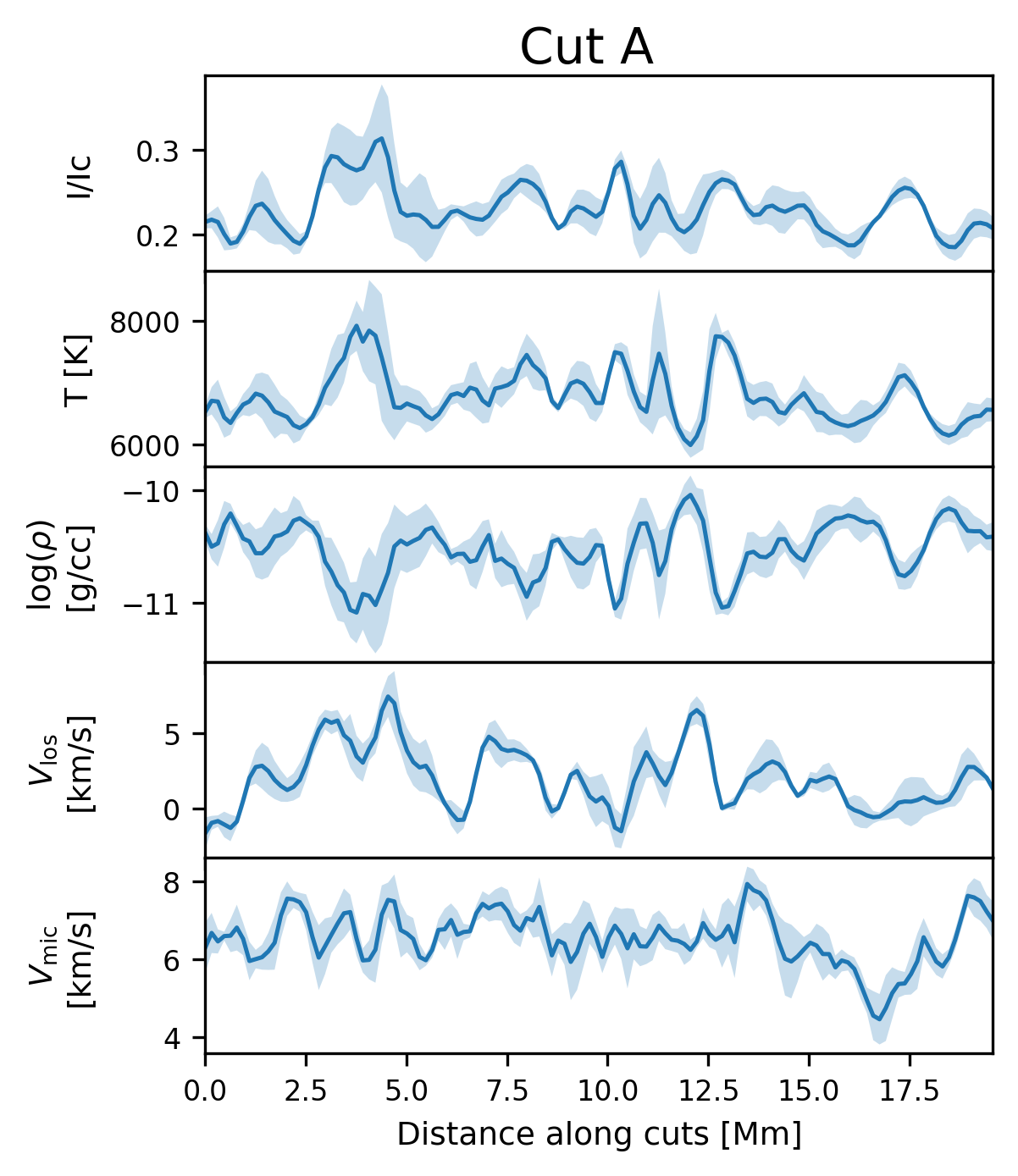}
\includegraphics[width=0.48\textwidth]{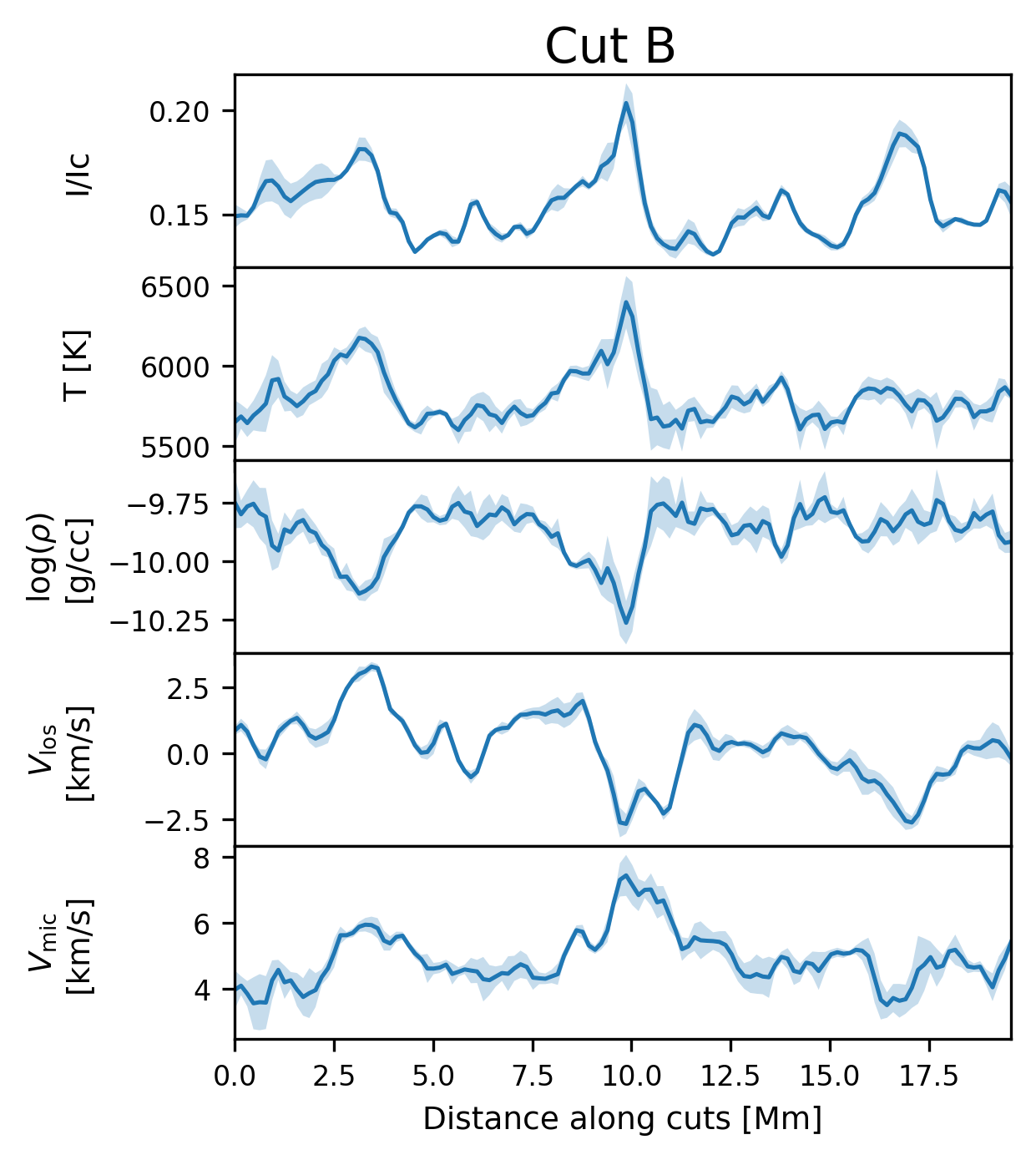}
\includegraphics[width=0.48\textwidth]{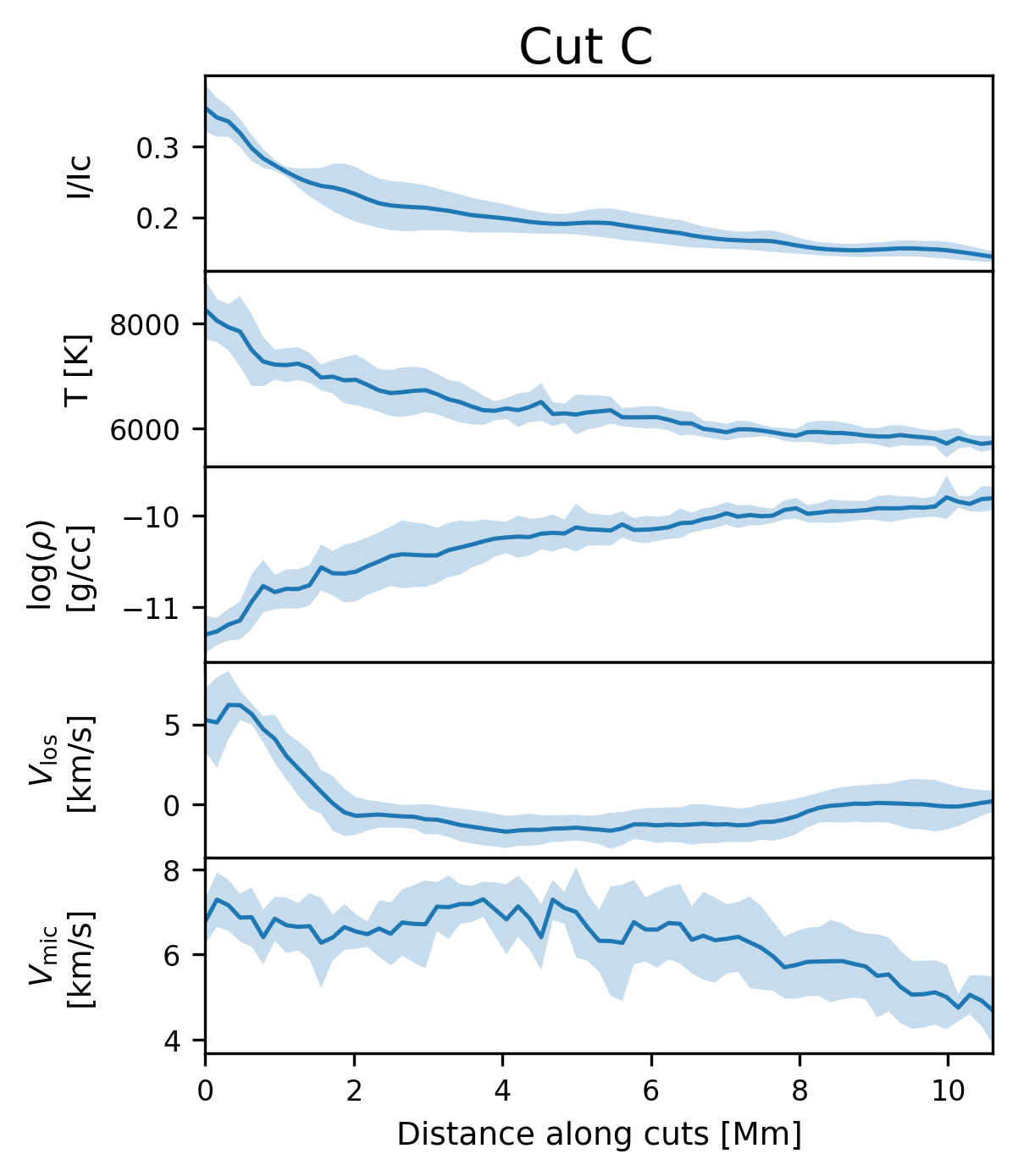}
\includegraphics[width=0.48\textwidth]{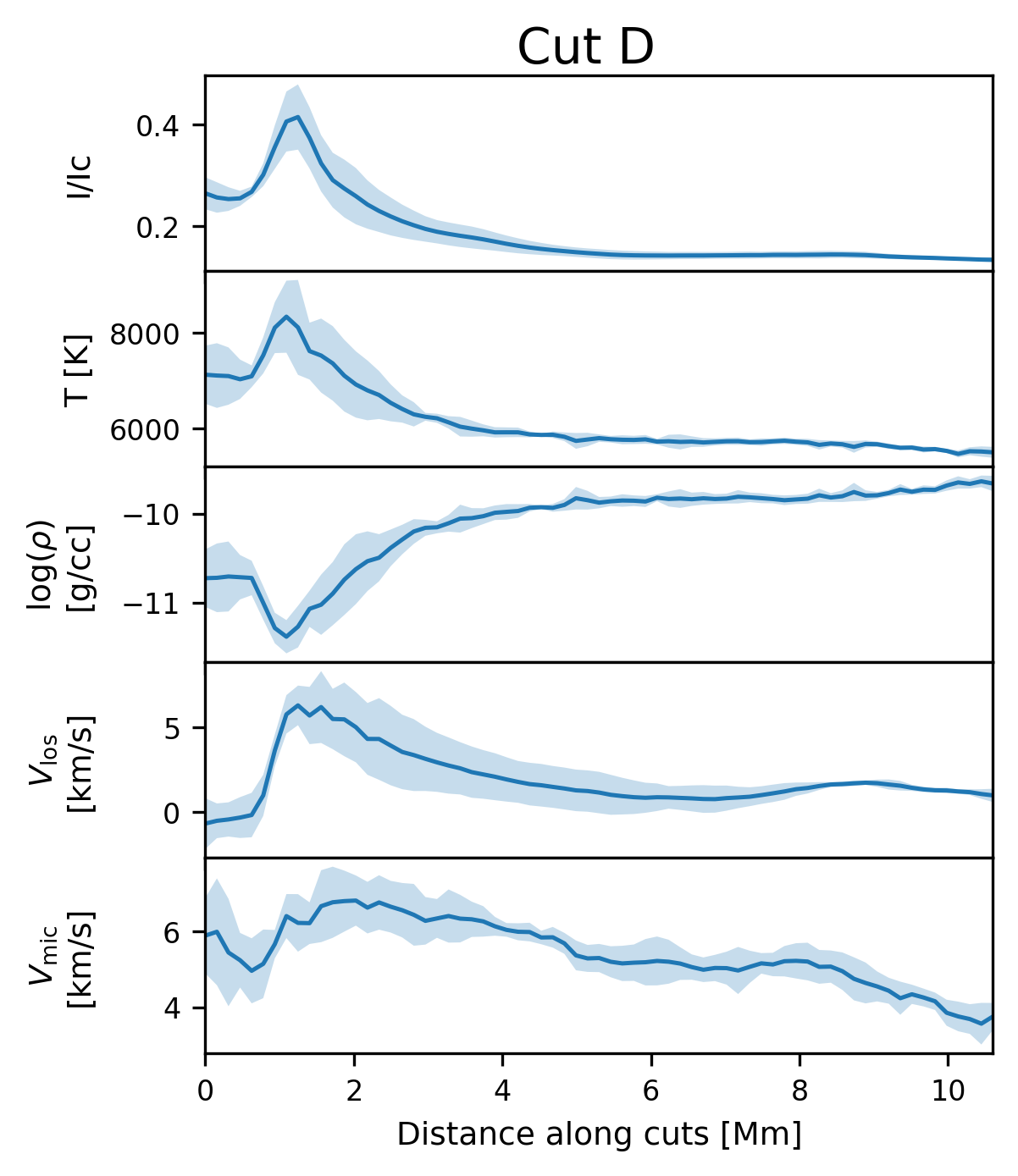}
\caption{The panels on the top row show the profile of line core intensity, temperature, density, Doppler velocity and microturbulent velocity across the fibrils, corresponding to the cuts  `A' (top left panel) and `B' (top right panel), as labeled in the Figure~\ref{fig:map_with_cuts}.Similarly, the bottom panels show the profile of these parameters along the fibrils, corresponding to the cuts `C' (bottom left panel) and `D' (bottom right panel). For all panels the blue shaded region depicts the standard deviation of the mean value (the solid curve).}
\label{fig:fibril_profile_alongacross_physical}
\end{figure*}

\subsubsection{Thermodynamic profile along and across fibrils}
Figure \ref{fig:map_with_cuts} presents a line-core intensity map overlaid with four diagnostic slits. Slits `A' and `B' are oriented perpendicular to the fibril axis (across-cut), while slits `C' and `D' follow the fibril length (along-cut). We extracted five key physical parameters along these trajectories: line-core intensity, temperature, gas density, Doppler velocity ($V_{\mathrm{los}}$), and microturbulent velocity ($V_{\mathrm{mic}}$). To ensure statistical robustness, each slit was sampled with a transverse width of 10 pixels. From this sample, we calculated both the mean spatial profile and the standard deviation at each point along the cut. These results are displayed in Figure \ref{fig:fibril_profile_alongacross_physical}, where the solid curves represent the mean values and the blue shaded regions denote the $\pm 1\sigma$ spatial dispersion.

\textit{Variation Across the Fibrils}
The variation of parameters across the fibrils is shown in the top row of Figure~\ref{fig:fibril_profile_alongacross_physical}. These curves highlight the sharp thermal and kinematic contrast between the fibril and inter-fibril region as well as correlation between these parameters.

\textbf{Slit `A' Profile:} 
\begin{enumerate}
    \item The temperature, density and relative line core intensity shows good spatial correspondence between each other. Peaks in temperature often coincides with peaks in core intensity and dips in density.
    \item The Doppler velocities across these cuts are predominantly positive (redshifts), reaching approximately 5\,km\,s$^{-1}$ near the fibril centres, indicating downflowing plasma. 
    \item Microturbulent velocities are high $\sim 4$--$9$~km\,s$^{-1}$ and do not show spatial correspondence with temperature.
\end{enumerate}  

\textbf{Slit `B' Profile:} 
\begin{enumerate}
    \item The temperature is on average lower by $\sim$1000K as compared to slit `A' and shows similar correspondence to density and line core intensity.
    \item The density is higher by an order of magnitude as compared to slit `A'.  
    \item The Doppler velocities are in general lower compared to slit `A' with few pronounced blue-shifted fibrils around 10 and 17 Mm mark on the X-axis. 
    \item Microturbulent velocities are slightly lower $\sim 4$--$6$~km\,s$^{-1}$, with the exception of a peak around 10 Mm mark on X-axis.
\end{enumerate}  

Both slit `A' and `B' show strong variation in temperature and density across fibrils, however, slit `A' shows much stronger gradients over a unit distance. This variation of `A' and `B' suggests that fibrils are thermally insulated from their surroundings, i.e., thermal conduction across the magnetic field lines is very low, and helps maintain such sharp gradients. Further, the steep gradient in density at the fibril edge suggests a magnetostatic pressure balance. To maintain the structure, the lower temperature inside the fibril is compensated by higher density and likely a stronger internal magnetic pressure to keep balance with surrounding hotter and less dense plasma.

\textit{Variation Along the Fibrils}
The variation of parameters along the length of fibrils is shown in the bottom row of Figure~\ref{fig:fibril_profile_alongacross_physical}. These curves highlight the variation of thermal and kinematic properties between the fibril footpoint and its main body.

\textbf{Slit `C' Profile:} 
\begin{enumerate}
    \item The temperature, density and relative line core intensity shows good spatial correspondence between each other. High temperature and line core intensity near the footpoint which decreases as one moves along the main body of fibril.
    \item The Doppler velocity near footpoint is strongly positive (5\,km\,s$^{-1}$ redshifts) decreasing in magnitude and changing sign (blueshift) as one moves along the fibril axis. Projection along line-of-sight may play a role in these variations as the plasma flow is presumably along the fibril which becomes predominantly horizontal near its main body.  
    \item Microturbulent velocity also seems high $\sim$7~km\,s$^{-1}$ near footpoint which decreases to $\sim$4~km\,s$^{-1}$ near its main body.  
\end{enumerate}  

\textbf{Slit `D' Profile:} 
\begin{enumerate}
    \item The slit `D' starts beyond footpoint and hence the temperature bump corresponding to footpoint is seen at around 1 Mm mark on the X-axis. The temperature near footpoint is  $\sim$8000K and decreases away from footpoint to $\sim$6000K after a distance of a few Mm. 
    \item The density at the footpoint is lower by an order of magnitude as compared to the main body.  
    \item The Doppler velocity is more than 5~km\,s$^{-1}$ near the footpoint and decreases towards the main body. 
    \item Microturbulent velocity profile starts high at $\sim7$~km\,s$^{-1}$ near the footpoint and  decreases towards main body. 
\end{enumerate}  

Both slit `C' and `D' show strong but smooth variation in temperature and density along the fibrils. Along the length of the fibril, the magnetic field $B$ is presumably parallel to our slits, as is shown in previous studies \citep{schad2013,aasensio2017}.  Therefore, the variations in temperature and density are likely balanced by gas pressure gradients rather than magnetic pressure jumps. This is why the gradients along the fibril are much smaller than the gradients across the fibril (slits `A' and `B').

\subsection{Discussion}
Our inversion results confirm that the chromospheric fibrils  are cooler, denser and dynamically distinct from their surrounding plage region.  The temperature of the fibrils near their footpoints  is typically $\sim$8000K and then rapidly drops to $\sim$6000 K along the fibrils length, i.e., after about 5Mm from its footpoint. Beyond this distance the temperature drops much more slowly to reach values below 6000~K. 
Our results suggest much larger temperature variation as compared to that reported by Kriginsky et~al.\ (2023), who found that fibril footpoints are about 300\,K hotter than their midpoints.  
While the exact values of temperature and density may be affected by the assumptions in the inversion process, such as hydrostatic equilibrium, the relative contrast between fibrils and their surroundings is likely robust.  The enhanced density within the fibrils supports the idea that they are mass-loaded structures, possibly due to shock-driven upflows or siphon flows along magnetic field lines.

Radiation–magnetohydrodynamic simulations predict that dark H\ensuremath{\alpha} and Ca \textsc{ii} fibrils correspond to regions of high mass density (e.g., see \cite{druett2022}), and our results support this interpretation. Enhanced microturbulent velocities within the fibrils further suggest that they host small–scale motions or waves, possibly related to the oscillatory motions and shock fronts observed in dynamic fibrils and spicules \citep{leenaarts2012}. Our study therefore reinforces the view that fibrils act as conduits for mass and energy transport in the chromosphere.  

Observations across the solar disk show that fibrils, mottles and spicules are manifestations of similar magnetically guided plasma structures rooted in network or plage magnetic concentrations (e.g., see \cite{dong2025}).  The density enhancements and downward flows in fibrils are explained with the idea that dense chromospheric material is lifted by magnetoacoustic shocks and subsequently drains back along inclined magnetic field lines, as in numerical models of spicule formation.  Moreover, we find that microturbulent velocities are significantly larger near the footpoints of fibrils which gradually decreases along their length, implying unresolved motions that may contribute to chromospheric heating.

Our inversion strategy based on k–means clustering and NLTE inversions proves to be very efficient.  By reducing the high–dimensional dataset to a small set of representative profiles and using their best–fit model atmospheres as initial guesses, we obtain smooth and physically plausible maps of thermodynamic parameters across the field of view.  A comparison with inversions starting from a single semi–empirical model shows that the k–means initialization yields lower chi–square values and avoids artificial small–scale fluctuations in the retrieved parameters (see Appendix~B).  While neural–network approaches have been explored as a fast alternative to traditional inversions \citep[e.g.][]{milic2020,chappell2022}, our two–step method is fast enough for the dataset analyzed here and offers the advantage of interpretable model atmospheres.

We note that our diagnostics are based solely on the Ca \textsc{ii} 854.2 nm line.  Although this line provides valuable information on the lower and middle chromosphere, its relatively broad formation height means that temperature gradients within fibrils may not be fully resolved.  Recent studies suggest that combining Ca \textsc{ii} lines with other diagnostics such as Mg \textsc{ii} h\&k or infrared triplet lines is necessary to disentangle temperature and microturbulence with high fidelity (e.g., see \cite{Kriginsky2023}).  Future observations with DKIST that sample multiple chromospheric lines simultaneously, together with advanced inversion techniques, will therefore be key to improving the diagnostic capability for chromospheric fibrils.

\section{Conclusions}
We have presented a k–means–assisted non–LTE inversion analysis of chromospheric fibrils observed with ViSP in the Ca \textsc{ii} 854.2 nm line.  The high spatial and spectral resolution of the dataset allowed us to apply unsupervised clustering to the Stokes~$I$ spectra, derive representative model atmospheres for each cluster and use these as initial models for pixel–by–pixel inversions.  Our method allowed us to retrieve smooth and physically plausible maps of temperature, density, velocity and microturbulence across the field of view.

Our inversions show that fibrils are cooler by 1–2\,kK and denser by an order of magnitude than the surrounding plage atmosphere in the optical depth range  $\log\tau = -4.5$ to $-5.5$.  The temperature varies along the fibril axis, with footpoints being hotter than midpoints by a couple of thousands of kelvin, much larger than estimated in recent high–resolution studies.  The Doppler maps reveal a predominance of downflows of a few km\,s$^{-1}$ within the fibrils, while microturbulent velocities are enhanced, indicating unresolved motions.  Cross–cut analyses confirm that the physical parameters vary strongly lateral to the fibril length, typically on a megameter scale.  

The clear separation of chromospheric structures achieved by our k–means clustering demonstrates the value of unsupervised machine–learning techniques as a pre–processing step for inversions.  Our two–step inversion procedure not only improves the fit quality compared with standard HSRA–initialised inversions but also enables an efficient exploration of the parameter space.  The results presented here provide observational constraints for theoretical models of fibril formation, heating and mass loading and highlight the need for multi–line diagnostics and numerical simulations to fully understand the complex dynamics of these ubiquitous solar features.

Future work will extend this analysis by including additional chromospheric lines, exploring time–dependent behaviour, and comparing with state–of–the–art 3D radiation–MHD simulations.  Such efforts will help to clarify the role of fibrils in chromospheric energy balance and their connection to other small–scale phenomena such as spicules and surges.

Finally, the biggest caveat of the present study is the assumption of hydrostatic equilibrium (HSE) which limits the accuracy of the inferred semi-empirical models. As demonstrated by \cite{henriques2020} the HSE assumption breaks down in the solar chromosphere due to neglect of dynamical effects and magnetic forces in the simplified momentum equation. Addressing this limitation in future studies will require going beyond HSE approximations, such as demonstrated by \cite{vincente2026} with their novel application of multi-line diagnostics and 3D magnetohydrostatic (MHS) equilibrium to constrain gas pressure and density.

\begin{acknowledgments}
The research reported herein is based in part on data collected with the
Daniel K. Inouye Solar Telescope (DKIST), a facility of the
National Solar Observatory (NSO). This work also utilizes SOLIS data obtained by the NSO Integrated Synoptic Program (NISP), managed by the National Solar Observatory, which is operated by the Association of Universities for Research in Astronomy (AURA), Inc. under a cooperative agreement with the National Science Foundation. DKIST is
located on land of spiritual and cultural significance to Native
Hawaiian people. The use of this important site to further
scientific knowledge is done so with appreciation and respect.
Any opinions, findings and conclusions or recommendations
expressed in this publication are those of the author(s) and do
not necessarily reflect the views of the National Science
Foundation or the Association of Universities for Research in
Astronomy, Inc. Author SG thanks Kevin Reardon, Han Uitenbroek, Rahul Yadav and Joao da Silva Santos for helpful discussions. Finally, we thank the anonymous referee for critical comments and suggestions that improved the presentation of the results.
\end{acknowledgments}

\facilities{DKIST(ViSP), SDO(AIA).}

\software{Astropy \citep{astropy2018}, SunPy \citep{sunpy_community2020}, NumPy \citep{numpyharris2020}, MatplotLib \citep{matplotHunter2007} and Scikit-learn \citep{scikit-learn}}

\appendix
\section{ViSP continuum normalization using FTS and SOLIS/VSM full-disk observations}
\begin{figure}[h]
\centering
\includegraphics[width=1.0\textwidth]{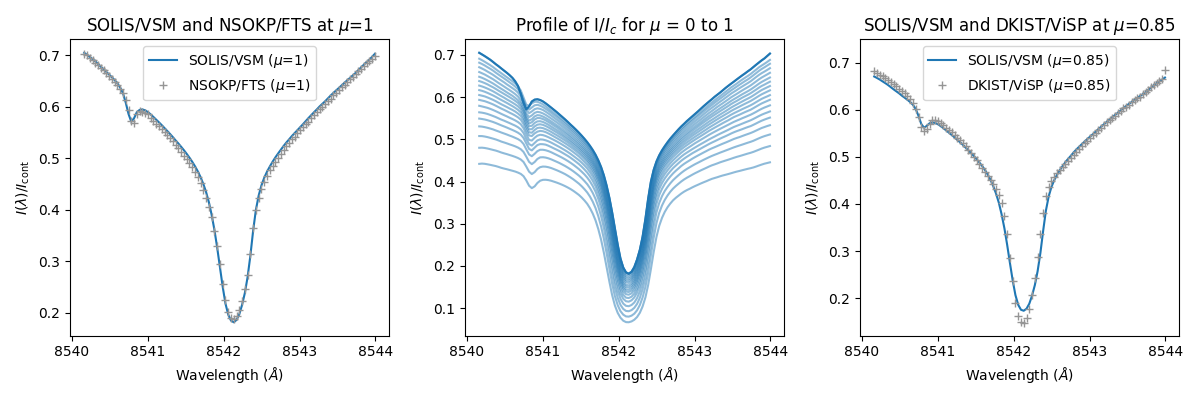}
\caption{Left panel shows the normalized NSOKP/FTS atlas spectrum at disk center (symbols) and the corresponding SOLIS/VSM spectrum (blue curve). Middle panel shows the quiet-sun spectra at various heliocentric angles. Right panel shows the quiet-sun ViSP/DKIST spectrum (symbols) normalized to the corresponding VSM/SOLIS spectrum (blue curve) at $\mu$=0.85. }
\label{fig:vsm_norm}
\end{figure}

Normalization of the ViSP spectra to the disk-center quiet-Sun intensity is performed using SOLIS/VSM data. The VSM is a full-disk spectropolarimeter designed for synoptic, long-term monitoring of the solar magnetic field \citep{keller1998}. It observes the photosphere and chromosphere using the Fe I 630 nm and Ca \textsc{ii} 854 nm spectral lines, respectively. A large spectrograph slit scans the solar image from pole to pole, producing spectral cubes of the format $I(x,y,\lambda)$ for each Stokes parameter. We use the Stokes I spectra for the continuum normalization of the ViSP spectra as follows:

(i) We fit the quiet-Sun disk-center Ca \textsc{ii} spectrum from the VSM to the FTS atlas spectrum and determine the continuum intensity, $I_{\text{cont}}(\mu=1)$. The left panel of figure~\ref{fig:vsm_norm} shows the SOLIS/VSM disk-center quiet-Sun spectrum (solid curve) fitted to the NSO/FTS atlas profile (+ symbols). The fitting yields the continuum intensity $I_{\text{cont}}$ and a stray-light factor for the VSM spectrum; the latter is subtracted from the SOLIS/VSM profiles before dividing by $I_{\text{cont}}$.

We use the Stokes I spectra for the continuum normalization of the ViSP spectra as follows: \
(i) We fit the quiet-sun disk-center Ca \textsc{ii} spectrum from the VSM to the FTS atlas spectrum and determine the continuum intensity, $I_{\text{cont}}(\mu=1)$. The left panel of figure~\ref{fig:vsm_norm} shows the SOLIS/VSM disk-center quiet-sun spectrum (solid curve) fitted to NSOKP/FTS atlas profile (+ symbols). The fitting yields the continuum intensity $I_{\text{cont}}$ and a stray light factor for VSM spectrum; the latter is subtracted from the SOLIS/VSM profiles before dividing by $I_{\text{cont}}$.

(ii) We normalize the spectra at each pixel of the VSM data-cube to obtain $I$($x$,$y$,$\lambda$)/$I_{\text{cont}}$, which is then used to derive mean quiet-sun spectrum at various $\mu$ positions, $I$($\mu$,$\lambda$)/$I_{\text{cont}}$.
The middle panel of figure~\ref{fig:vsm_norm} shows the normalized quiet sun spectra from SOLIS/VSM at various heliocentric angles, i.e., from $\mu$=1 to 0.1 in steps of 0.05, in the order of decreasing relative intensity, respectively. It may be noted that Doppler shift due to solar rotation has been removed from these spectra.

(iii) We fit the quiet-sun spectrum from ViSP observations to the VSM quiet-sun spectrum corresponding to $\mu$=0.85 (position of our ViSP scan). The right panel of figure~\ref{fig:vsm_norm} shows quiet-sun profile of ViSP observations fitted to VSM spectra. This fitting yields continuum normalization for the ViSP data. 

\section{Comparison of Chi square between K-means and HSRA model initializations}
\begin{figure}[h]
\centering
\includegraphics[width=0.75\textwidth]{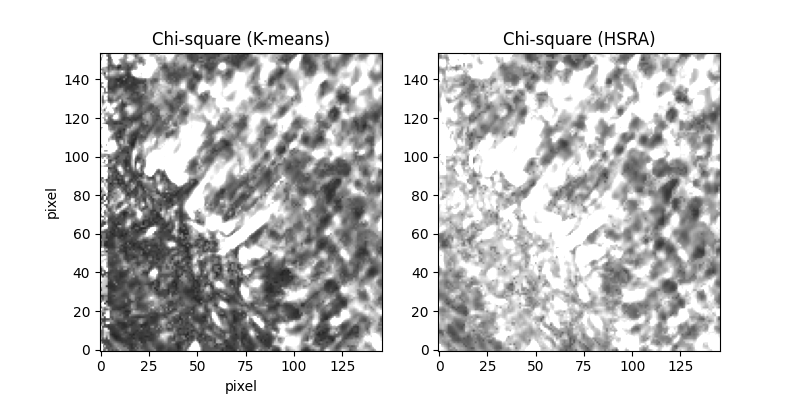}
\caption{Map fo chi-square statistic when the inversions are initialized with K-means approach (left panel), and when they are initialized with HSRA model for each pixel (right panel). }
\label{fig:chisq_map}
\end{figure}

A map of chi-square statistic is shown in the upper panels of figure~\ref{fig:chisq_map} for comparison between the K-means initialized versus HSRA initialized inversions of the spectra. It may be seen that the chi-square, a goodness-of-fit statistic, is lower (darker areas) in the K-means initialized inversions as compared to HSRA initialized ones. 

\begin{figure}[h]
\centering
\includegraphics[width=0.75\textwidth]{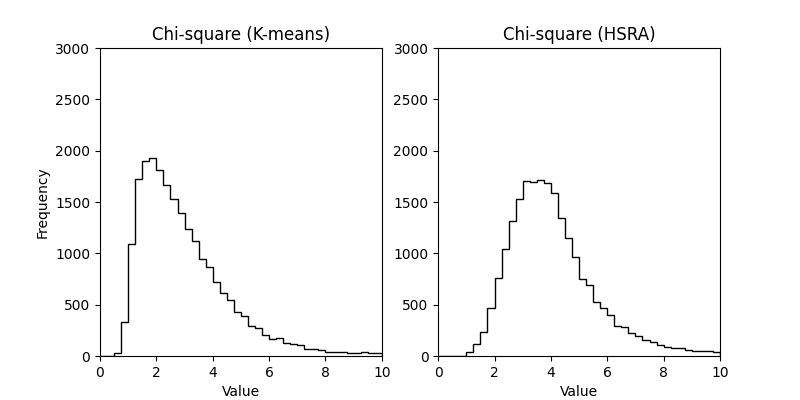}
\caption{The histogram of chi-square statistic when the inversions are initialized using the K-means approach and when they are initialized with HSRA model for each pixel (right panel).}
\label{fig:chisq_hist}
\end{figure}

The histograms corresponding to these chi-square maps is shown in the figure~\ref{fig:chisq_hist}. Again we can see that the chi-square distribution is narrower and peaks at lower value for the K-means initialization as compared to the HSRA initialization.  

\bibliography{sample631}{}
\bibliographystyle{aasjournal}

\end{document}